# Multiline Orthogonal Scanning Temporal Focusing (mosTF) Microscopy for Scattering Reduction in High-speed *in vivo* Brain Imaging

Yi Xue[1,2,8], Josiah R. Boivin[3], Dushan N. Wadduwage[2,4,7], Jong Kang Park[4], Elly Nedivi[3,5,6] and Peter T.C. So[1,2,4,*]

**Abstract**

Temporal focusing two-photon microscopy enables high resolution imaging of fine structures *in vivo* over a large volume. A limitation of temporal focusing is that signal-to-background ratio and resolution degrade rapidly with increasing imaging depth. This degradation originates from the scattered emission photons are widely distributed resulting in a strong background. We have developed Multiline Orthogonal Scanning Temporal Focusing (mosTF) microscopy that overcomes this problem. mosTF captures a sequence of images at each scan location of the excitation line, followed by a reconstruction algorithm reassigns scattered photons back to the correct scan position. We demonstrate mosTF by acquiring mice neuronal images *in vivo*. Our results show remarkably improvements with mosTF for *in vivo* brain imaging while maintaining its speed advantage.

**Introduction**

Two-photon laser scanning microscope (TPLSM) has been widely used for *in vivo* imaging of neuronal dendritic structures[1-2]. However, for TPLSM, high resolution imaging of the fast remodeling dynamics of fine neuronal structures, such as spines, remains a challenge[3-4]. As an alternative to TPLSM, line-scanning temporal focusing microscopy (lineTF) has been used for fast two-photon imaging in mouse brain[5-6]. lineTF focuses light spatially along one axis and modulates pulse dispersion along the other axis using the principle of "temporal focusing"[7-9]. Instead of point-scanning, line scanning improved imaging speed by over an order of magnitude[5]. Even though lineTF has the same excitation point spread function as TPLSM, the theoretical resolution has never been realized *in vivo*. Because lineTF requires a camera for whole field image detection, it is severely influenced by tissue scattering of the emission photons[10]. Scattered emission photons are distributed over a broad range of pixels adjacent to the scan locations, decreasing signal intensity and increasing background noise. Thus, the signal-to-noise ratio (SNR) and signal-to-background ratio (SBR) of lineTF are greatly degraded compared to TPLSM.

[1]Dept. of Mechanical Engineering, [2]Laser Biomedical Research Center, [3]Picower Institute, [4]Dept. of Biological Engineering, [5]Dept. of Biology, [6]Dept. of Brain and Cognitive Sciences, Massachusetts Institute of Technology 77 Massachusetts Ave., Cambridge MA 02139. [7]Center for Advanced Imaging, Faculty of Arts and Sciences, Harvard University, Cambridge, MA 02138. [8]Dept. of Electrical Engineering and Computer Sciences, University of California, Berkeley, CA 94720, USA. *Author e-mail address: ptso@mit.edu

Many methods have been developed to image or focus light through turbid media. For example, optical phase-conjugation has been used to enhance light transmission through rabbit ear[11] and mouse skin[12]. This method collects transmission photons and assumes the scattering process is time-reversible, which is not feasible for mouse brain imaging. Other methods collect back scattered photons for turbidity suppression but require iterative illumination[13] or angular scanning[14] to record tens to hundreds of images for reconstruction, but their utility in living tissues is limited by the millisecond scale speckle decorrelation times. In addition, these methods [11-14] have succeeded in focusing light or imaging low resolution targets through living tissues but have not been applied to image tissue internal structures at diffraction-limit level resolution.

Structured illumination combined with image processing is used to reduce out-of-focus scattering for temporal focusing microscopy[5,15-18]. With only one structured illuminated image[5,15-16], only the ballistic photons are utilized for image reconstruction; the scattered emission photons are simply discarded by low-pass filtering. This method improves image contrast but cannot recover information carried by the scattered emission photons. With multiple structured illuminated images[17-18], typically hundreds to thousands, one can recover an image of the object at the resolution of the excitation independent of image blurring by the scattering of the emission photons. While these methods can effectively reduce aberration and scattering, imaging speed is compromised to recover accurate image features.

Here, we developed Multiline Orthogonal Scanning Temporal Focusing (mosTF) microscopy to overcome the emission photon scattering problem without any prior information of the turbid tissue, while mostly maintaining the speed advantage of temporal focusing. Several excitation lines at either horizontal or vertical orientations are scanned to cover the whole field of view (FOV) (Fig. 1a, see setup details in Methods). It is important to note that at each scan location, a fast EMCCD camera captures an intermediate image. Therefore, the effect of emission photon scattering can be mostly compensated similarly to TPLSM, where scattered photons can be reassigned back to the excitation locations, allowing the reconstruction of a final image at high SBR and resolution (Fig. 1b, see reconstruction details in Methods). However, unlike for point scanning, line scanning results in additional ambiguity in photon reassignment and image anisotropy. Reconstructing two orthogonally scanned intermediate images mostly recovers an isotropic image (Fig. S1). After reconstruction, mosTF reveals small structures that would be invisible using lineTF by significantly improved image SBR. We demonstrate the performance of mosTF by *in vivo* imaging of mice neurons under anesthesia. mosTF is able to image dendrites and dendritic spines over a large FOV in cortical layer 2/3 (about 170 µm deep inside the brain) with four times higher SBR than lineTF and eight times faster imaging speed than TPLSM. mosTF could be a useful technique for real-time visualization of synaptic dynamics, which is a fundamental feature of brain development and plasticity[2-4, 19-20].

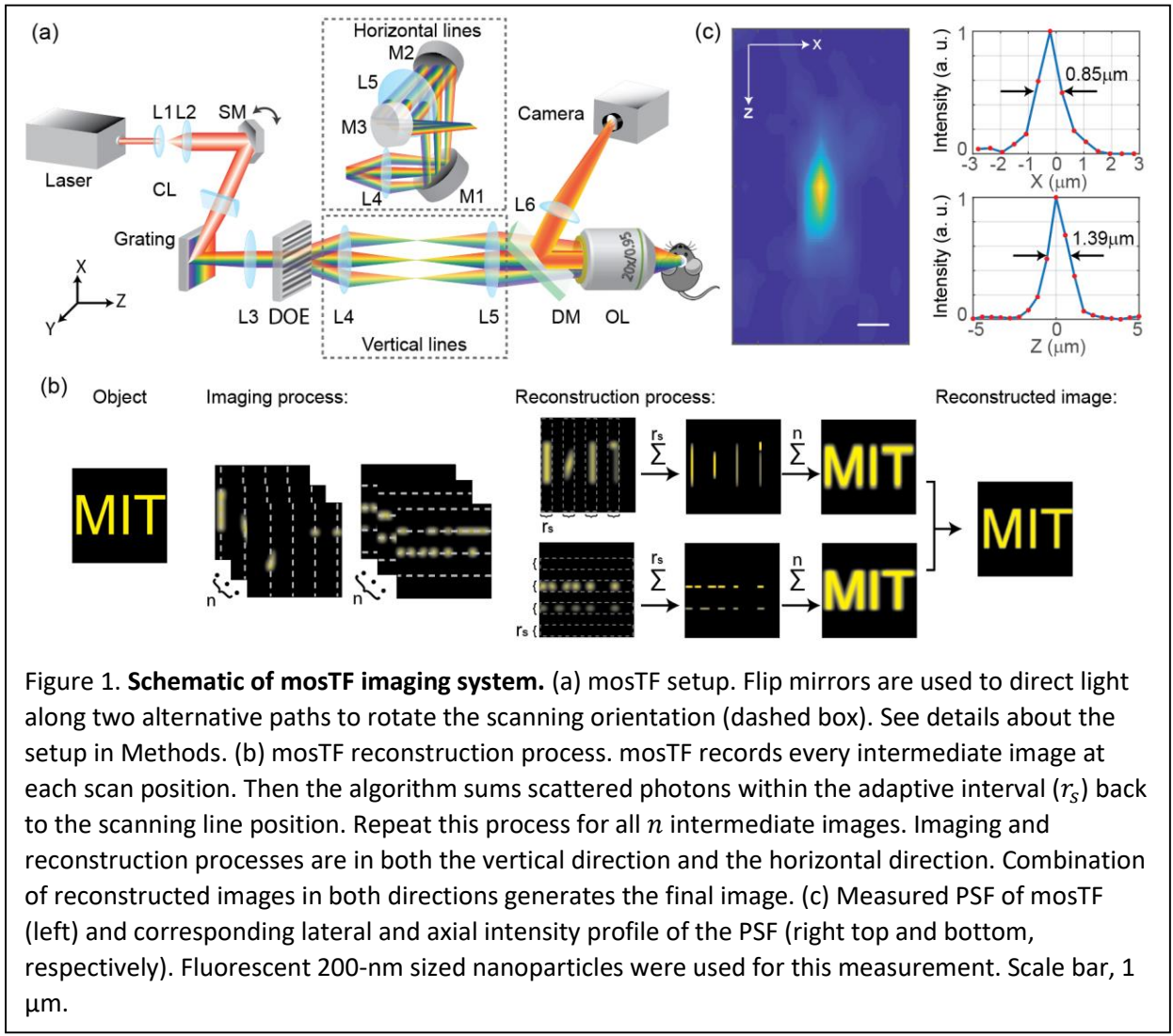

Figure 1. **Schematic of mosTF imaging system.** (a) mosTF setup. Flip mirrors are used to direct light along two alternative paths to rotate the scanning orientation (dashed box). See details about the setup in Methods. (b) mosTF reconstruction process. mosTF records every intermediate image at each scan position. Then the algorithm sums scattered photons within the adaptive interval ($r_s$) back to the scanning line position. Repeat this process for all $n$ intermediate images. Imaging and reconstruction processes are in both the vertical direction and the horizontal direction. Combination of reconstructed images in both directions generates the final image. (c) Measured PSF of mosTF (left) and corresponding lateral and axial intensity profile of the PSF (right top and bottom, respectively). Fluorescent 200-nm sized nanoparticles were used for this measurement. Scale bar, 1 µm.

## Results

### 1. Imaging through scattering media

      As a first experiment, we imaged the same area of fluorescent beads using water (minimal scattering, Fig. 2) and then 0.2% intralipid solution (tissue-level scattering, Fig. 3) as the objective immersion media. The mosTF images were generated by the algorithm described in the Methods section. The images without reconstruction, that is, lineTF images, were generated by directly summing the intermediate images together without photon reassignment.

      For the control experiment using water as the immersion media, we first quantified the image quality by evaluating the distribution of the beads' intensity, SBR and Mean-Square-Error (MSE) in the large FOV (Fig. 2 a1-b1). The distributions of the beads' intensity were nearly the

same for the mosTF and lineTF images (Fig. 2c). The SBR for the mosTF was improved three fold after reconstruction (Fig. 2d). Even with minimal scattering medium, photons from a point source were still detected by multiple pixels. The reconstruction process was similar to binning. MSE is calculated with respect to mosTF image, which shows the structural similarity between two images. MSE of lineTF image is 1.23, resulting from noise in the image. The zoomed-in view of individual beads (Fig. 2 a2-b2, e) shows an improvement of signal intensity. The reconstructed PSF is isotropic (Fig. 2f) because of the orthogonal scanning and combination. The intermediate images with horizontal and vertical scanning, as well as the corresponding Fourier domain images, are shown in Supplementary Figure 1. In summary, mosTF slightly improves image SBR compared to lineTF when water is used as a minimal scattering immersion media.

We subsequently compared the bead images taken by mosTF and lineTF with the 0.2% intralipid solution to mimic the effects of tissue scattering (Fig. 3). The statistical analysis of the full FOV shows an obvious advantage for mosTF (Fig. 3 a1-b1, c-d). Imaging through 2 mm 0.2% intralipid solution (about 3.2 $l_s^{em}$ where $l_s^{em}$ is the mean free path length of the emission photons; see calculations in Methods), lineTF can barely detect signal from the beads, while mosTF is able to identify individual single beads (Fig. 3c). Compared to lineTF, mosTF improves SBR by 36 times. The SBR of mosTF images through turbid media (70.45) is even higher than lineTF images without turbid media (47.61). Referring to the mosTF image with water immersion, the MSE of mosTF with lipid immersion is 3.53 and the MSE of lineTF with lipid immersion is 9.18. The MSE results elucidate mosTF image captured similar structural information with images taken through water immersion, while lineTF image lost structural information. In addition to the analysis of the full FOV, we also analyzed individual beads in the zoomed-in view (Fig. 3 a2, b2, e-f). The intensity profile shows that lineTF cannot distinguish signal and noise in this case, while mosTF still clearly resolves two adjacent beads (Fig. 3e). The PSF measurements along three different directions (x, y, and diagonal) are nearly identical, indicating the reconstructed PSF is isotropic. Thus, mosTF provides a significant enhancement in SBR and collected most of structural information when imaging through turbid media.

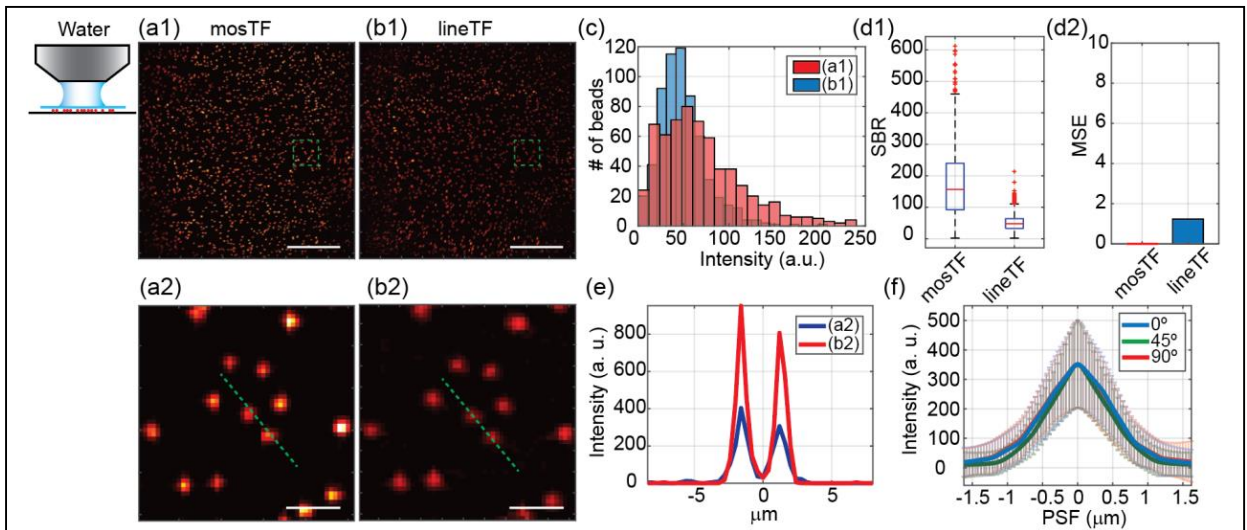

Figure 2. Images of 200 nm beads through water taken by mosTF and lineTF. The working distance of the objective is 2 mm. The full FOV is 205 x 205 µm². (a1) mosTF image of the full FOV. (a2) Zoomed-in view of the area labeled by a green box in (a1). (b1) LineTF image of the full FOV. (b2) Zoomed-in view of the area labeled by a green box in (b1). (c) Intensity distribution of beads in (a1, red) and (b1, blue). (d1) SBR comparison ($n$ = 619). SBR of mosTF: 156.66; SBR of lineTF: 47.62. (d2) MSE comparison. mosTF image works as a reference, so MSE of mosTF is zero. MSE of lineTF is 1.23. (e) Intensity profile of the cross-section of the two beads in (a2) and (b2), labeled by green dashed line. PSF profile along x- (blue), y- (red) and diagonal (green) direction ($n$ = 295). FWHM$_x$: 1.15 µm; FWHM$_y$: 1.15 µm; FWHM$_{diag}$: 1.00 µm. Scale bar in (a1, b1): 50 µm. Scale bar in (a2, b2): 5 µm.

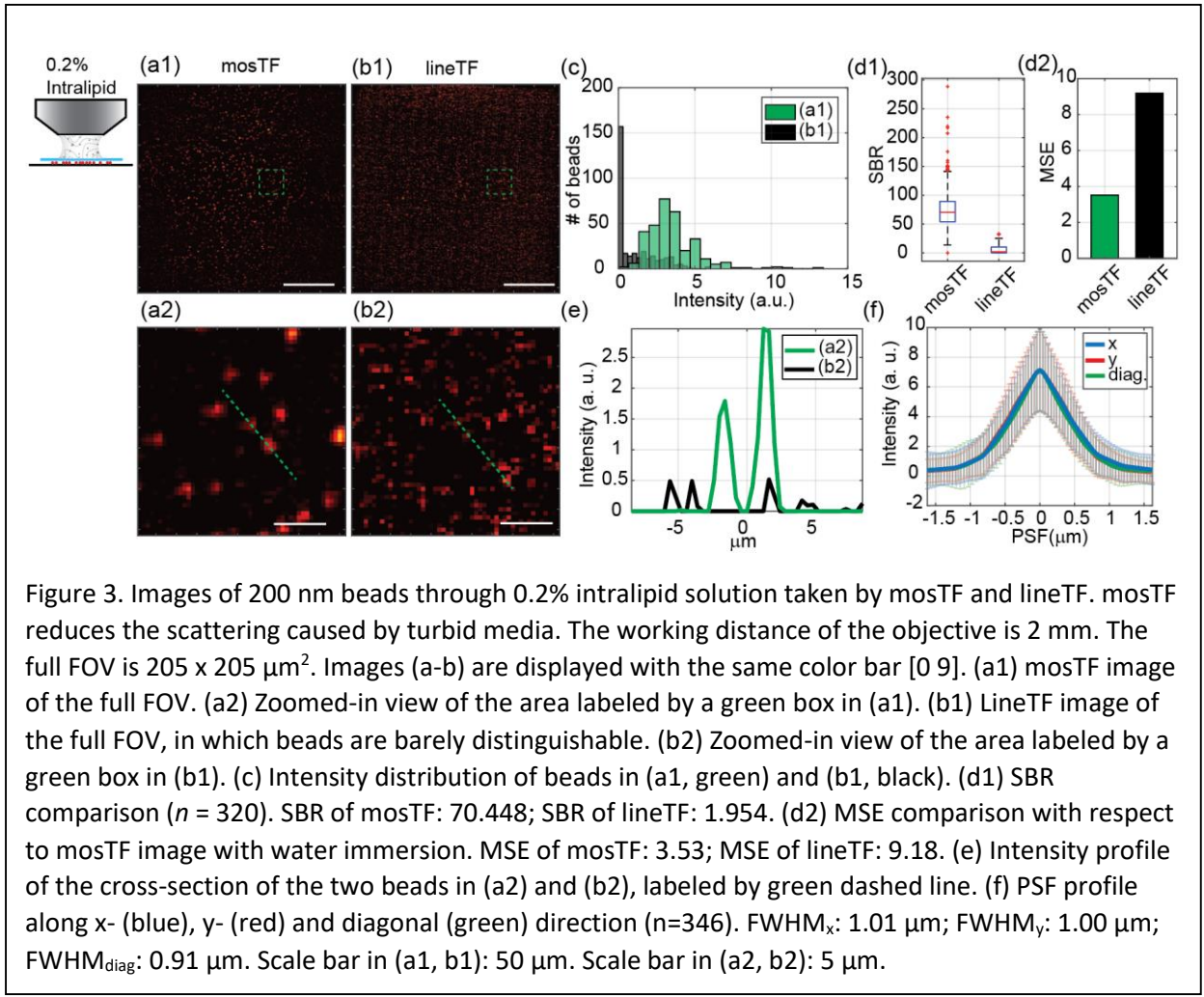

Figure 3. Images of 200 nm beads through 0.2% intralipid solution taken by mosTF and lineTF. mosTF reduces the scattering caused by turbid media. The working distance of the objective is 2 mm. The full FOV is 205 x 205 µm². Images (a-b) are displayed with the same color bar [0 9]. (a1) mosTF image of the full FOV. (a2) Zoomed-in view of the area labeled by a green box in (a1). (b1) LineTF image of the full FOV, in which beads are barely distinguishable. (b2) Zoomed-in view of the area labeled by a green box in (b1). (c) Intensity distribution of beads in (a1, green) and (b1, black). (d1) SBR comparison ($n$ = 320). SBR of mosTF: 70.448; SBR of lineTF: 1.954. (d2) MSE comparison with respect to mosTF image with water immersion. MSE of mosTF: 3.53; MSE of lineTF: 9.18. (e) Intensity profile of the cross-section of the two beads in (a2) and (b2), labeled by green dashed line. (f) PSF profile along x- (blue), y- (red) and diagonal (green) direction (n=346). $FWHM_x$: 1.01 µm; $FWHM_y$: 1.00 µm; $FWHM_{diag}$: 0.91 µm. Scale bar in (a1, b1): 50 µm. Scale bar in (a2, b2): 5 µm.

## 2. *In vivo* brain imaging of mice under anesthesia

We tested the performance of mosTF microscopy for *in vivo* imaging in the brain of an anesthetized mouse through a cranial window (see animal procedure in Methods). The sample is highly heterogeneous because of layers of different materials: cover glass, meninges, brain matter and blood vessels. Further, biological processes such as muscle contractions from breathing and heart beat can introduce unpredictable tissue movement[11-12]. However, since mosTF is able to image through turbid media without prior measurement, it can image through dynamic turbid media as long as the dynamics is slower than time required to take the two orthogonal scans. Also, the reconstruction process of mosTF treats every diffraction-limited region individually so that it is robust to spatially varied turbid media.

We imaged layer 2/3 pyramidal neurons labeled with eYFP as cell fill in an anesthetized mouse with mosTF, lineTF, and then TPLSM to generate a "ground truth" image. One focal plane, containing both soma and dendritic spines that carry excitatory synapses, located 170 µm deep inside the mouse brain (about 3 $l_s^{em}$, calculations are in Methods) is shown as a demonstration. The pixel size is 0.4 µm and the FOV is about 205 x 205 µm² (Fig. 4 a1-b1). The

TPLSM image has the same FOV, and the pixel size is 0.25 μm. The signal from dendritic spines is about 10-20 photons, while the soma is about 500 photons (measured from the TPLSM image). In the lineTF image, scattered photons from the soma are brighter than the nearby spines. As a result, the signal from the spines is indistinguishable from background noise. In comparison, mosTF reassigned the scattered photons back to their origin so that the spines on the dendritic shaft are clearly visible (Fig. 4 a2-b2). To quantitatively show the advantage of mosTF, the average SBR were measured from 49 individual locations, including soma, dendrites and spines. The SBR of mosTF (49.56) is nearly four times higher than the SBR of lineTF (13.08) (Fig. 4 d1). To evaluate structural similarity of fine structures, MSE of the mosTF and lineTF images were also calculated with respect to the TPLSM image (Fig. 4 d2) in the region of dendrites (Fig. 4 a3, b3, c3). mosTF image has better MSE (1.10) compared to lineTF image (2.53), which means mosTF recovered most of fine structures, such as dendritic and spines, as well as reduced background noise compared to lineTF. The SBR and MSE results show that mosTF effectively reassigned scattering photons to the correct pixels. In addition, orthogonal scanning removed scattering and uniformly extended frequency coverage in both horizontal (green dashed line, Fig 4 a2-b2) and vertical directions (green dashed line, Fig 4 a3-b3). The representative intensity profile is shown in both the x and y directions (Fig. 4e). More importantly, the intensity profiles of mosTF and lineTF show that mosTF not only increases the signal intensity but also reduces the background noise. Thus, mosTF's effective photon reassignment leads to an improvement in SBR and structural information fidelity. Consequently, mosTF is able to resolve small structures, such as dendritic spines, within the large FOV *in vivo*.

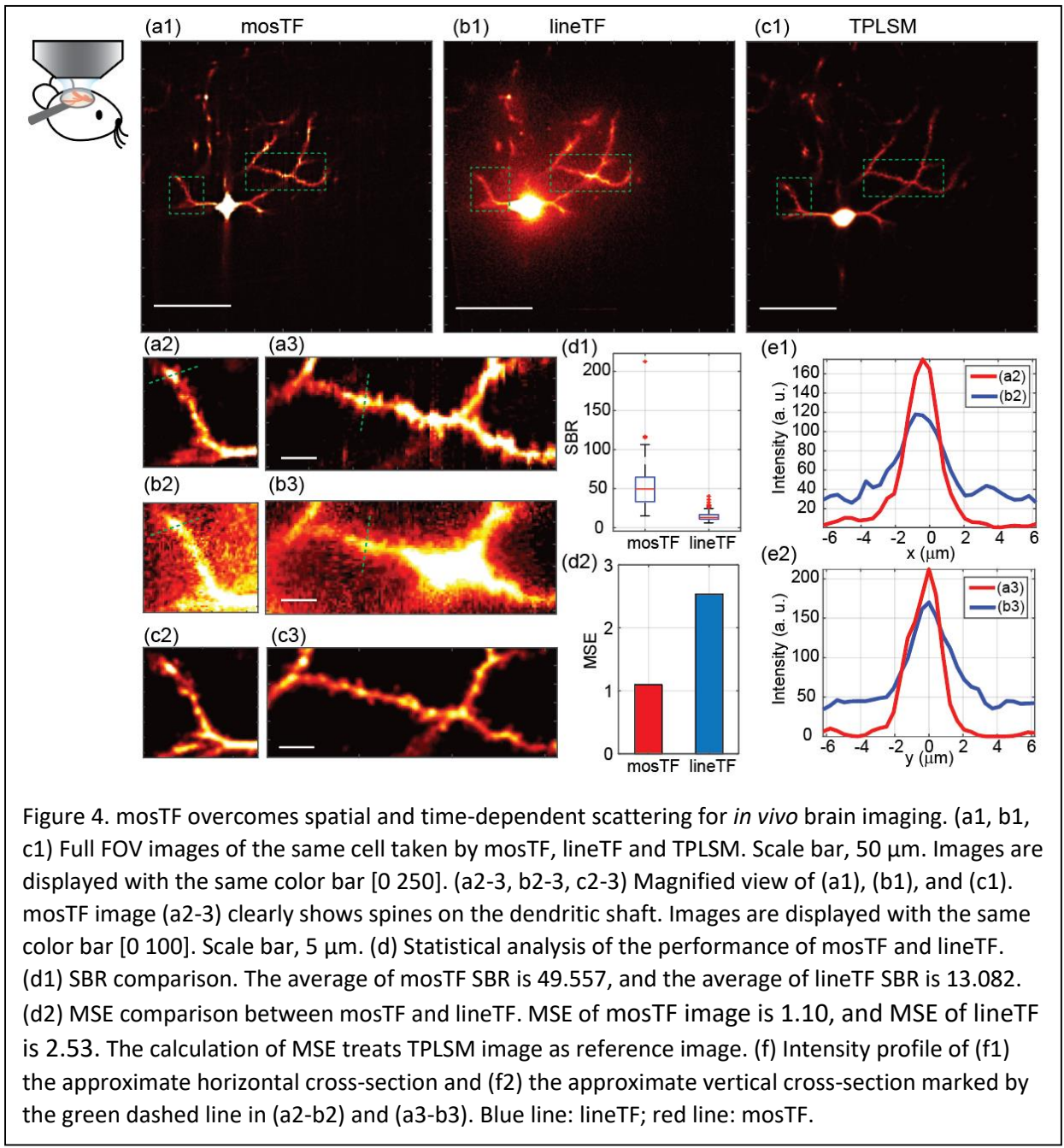

Figure 4. mosTF overcomes spatial and time-dependent scattering for *in vivo* brain imaging. (a1, b1, c1) Full FOV images of the same cell taken by mosTF, lineTF and TPLSM. Scale bar, 50 μm. Images are displayed with the same color bar [0 250]. (a2-3, b2-3, c2-3) Magnified view of (a1), (b1), and (c1). mosTF image (a2-3) clearly shows spines on the dendritic shaft. Images are displayed with the same color bar [0 100]. Scale bar, 5 μm. (d) Statistical analysis of the performance of mosTF and lineTF. (d1) SBR comparison. The average of mosTF SBR is 49.557, and the average of lineTF SBR is 13.082. (d2) MSE comparison between mosTF and lineTF. MSE of mosTF image is 1.10, and MSE of lineTF is 2.53. The calculation of MSE treats TPLSM image as reference image. (f) Intensity profile of (f1) the approximate horizontal cross-section and (f2) the approximate vertical cross-section marked by the green dashed line in (a2-b2) and (a3-b3). Blue line: lineTF; red line: mosTF.

**Discussion**

By reassignment of scattered photons, mosTF remarkably improves SBR and preserves fine structures when imaging through turbid media compared to lineTF. mosTF is able to image eight times faster than TPLSM while keeping the same spatial resolution and high SBR (see calculations in Methods). We have demonstrated the effectiveness of mosTF by imaging 200 nm red fluorescent beads through 2 mm thick 0.2% intralipid solution. While the lineTF image cannot distinguish any beads, the mosTF image can distinguish individual beads with high SBR. We also show *in vivo* brain imaging in an anesthetized mouse, without any prior correction for scattering. mosTF can clearly identify dendritic spines, on the order of 1 micron in size across a 205 x 205 µm$^2$ FOV, while in lineTF images these structures are mostly masked by scattered photons from adjacent bright structures. mosTF has the advantage of SBR and MSE over lineTF for *in vivo* imaging. In addition, mosTF achieves isotropic scattering reduction by orthogonal scanning.

Compared to other post-processing methods to reduce scattering[5, 15-16], mosTF reassigns the scattered photons back to their original position rather than simply rejecting them. Previous work using structured or patterned illumination[5, 15-16] takes multiple images under different illumination. Only in-focus information encoded by the pattern is preserved, while out-of-focus scattered photons are rejected by the post-processing algorithm. These previous methods can only increase SBR but not SNR because the scattered emission photons are lost from the signal but added to the in-focus background. However, mosTF not only reduces background noise caused by scattered photons but also increases signal intensity by reassigning these photons. This ability to reassign photons accounts for the large improvement in SBR and structural information maintainability seen with mosTF, allowing visualization of small, weakly fluorescent structures that are invisible with lineTF.

The imaging speed of mosTF can potentially be further improved in two ways: a camera with faster imaging speed and a laser with shorter pulse duration. As discussed in Methods, with a faster and lower readout noise camera, the theoretical speed improvement of mosTF using our 4-line scanning strategy would be approximately 40-fold compared to TPLSM, while we achieved eight-fold speed improvement limited by the camera used. If imaging brighter fluorescence samples, such as somas instead of spines, the imaging speed could be even faster to achieve sufficient SNR (see calculations in Methods). The second factor to improve mosTF imaging speed is the number of parallel lines. We can shorten the pulse duration of the laser from 200 fs to 50 fs by dispersion compensation. With four times shorter pulse duration, only half of the power per line can achieve the same two-photon excitation efficiency[21]. That means the number of parallel lines could be doubled using the same total power that is limited by the tissue thermal damage threshold. These two factors could potentially improve imaging speed of mosTF further by a factor of 10 in total with upgrades to the camera and laser.

According to the theory (see reconstruction algorithms in Methods), one-dimensional reconstruction leads to anisotropic PSF of mosTF. As a proof of principle setup, we introduce orthogonal scanning that rotates the beam by $\pi/2$ with mirrors. The current PSF has a slight "cross" shape, but could be made more isotropic with more rotation angles. Beam rotation at arbitrary angle can be realized using a Dove prism with a pre-chirp modulation to keep the pulse duration short.

In summary, mosTF is a novel approach for fast two-photon imaging through scattering media that presents improvement of signal-to background ratio and structural information maintainability compared to line-scan temporal focusing microscopy, and a speed improvement over point-scan two-photon microscopy. mosTF can be used for *in vivo* imaging of dynamic events on small structures across a large field for view in scattering tissue such as the brain, even in the presence of motion artifacts.

**Methods**

**1. mosTF reconstruction algorithm**

lineTF in the presence of emission scattering can be readily modeled. mosTF applies several excitation lines scanned in parallel at either horizontal or vertical orientations combined via the reconstruction algorithm. Here, we focus on a 2D reconstruction. We assume that the excitation point spread function (PSF) is theoretical and is not affected by the specimen. This assumption is mostly valid within 200 µm deep inside the mouse brain, because the tissue scattering and aberration of infrared excitation light is minimal[22]. We treat the emission PSF as unknown and potentially spatially varied due to local tissue heterogeneity. Therefore, the general form of each intermediate image at a scanning location $(x_0, y_0)$ within the FOV ($N_x \times N_y$ pixels, $1 \leq x_0 \leq N_x, 1 \leq y_0 \leq N_y$) can be written as:

$$I_1(x, y; x_0, y_0) = O(x,y)[E(x,y; x_0, y_0) \otimes P_{ex}(x,y)]] \otimes P_{em}(x, y; x_0, y_0), \tag{1}$$

where $P_{ex}(x,y)$ is the excitation PSF, $P_{em}(x, y; x_0, y_0)$ is the emission PSF that can vary dependent on excitation location, $O(x, y)$ is the fluorophore distribution and $E(x, y)$ is the excitation line pattern. Without loss of generality, we first consider a single scanning line at position $x_0$. Thus, $E(x, y; x_0, y_0) = \delta(x - x_0)$. Eq. (1) can be written as:

$$\begin{aligned} I_1(x, y; x_0, y_0) &= O(x,y)[\delta(x - x_0)P_{ex}(x,y)] \otimes P_{em}(x, y; x_0, y_0) \\ &= O(x,y)P_{ex}(x - x_0, y) \otimes P_{em}(x, y; x_0, y_0) \\ &= \iint O(x', y')P_{ex}(x' - x_0, y')P_{em}(x - x', y - y'; x_0, y_0)dx'dy' \end{aligned} \tag{2}$$

Theoretically, emission fluorescence photons should be integrated across the whole FOV. However, integral of pixels where photon counts are below readout noise of the camera only adds readout noise rather than signal to the reconstructed image. To optimize the SNR, only pixels with intensity above readout noise were integrated. Here, we applied an adaptive interval of the integral $r_s(x_0, y_0)$, which varies spatially depending on the local intensity. We can model this integral process as

$$\begin{aligned} I_2(y; x_0, y_0) &= \int_{-r_s(x_0,y_0)}^{r_s(x_0,y_0)} I_1(x, y; x_0, y_0)dx \\ &= \iint O(x', y')P_{ex}(x' - x_0, y') \left[\int_{-r_s(x_0,y_0)}^{r_s(x_0,y_0)} P_{em}(x - x', y - y'; x_0, y_0)dx\right]dx'dy' \end{aligned} \tag{3}$$

To elucidate the reconstruction process further, we define the 1D integral of $P_{em}$ is $\tilde{P}_{em}$. Eq. 3 can be rewritten as

$$I_2(y; x_0, y_0) = \iint O(x', y')P_{ex}(x' - x_0, y')\tilde{P}_{em}(x', y - y'; x_0, y_0)dx'dy'$$

(4)

To calculate the reconstructed PSF , $O(x,y) = \delta(x,y)$. Eq. 4 can be rewritten as

$$I^x_{PSF}(y; x_0, y_0) = \iint \delta(x', y') P_{ex}(x' - x_0, y') \tilde{P}_{em}(x', y - y'; x_0, y_0) dx' dy' \qquad (5)$$
$$= P_{ex}(-x_0, 0) \tilde{P}_{em}(0, y; x_0, y_0).$$

Thus, the reconstruction PSF at location $(x_0, y_0)$ is the product of excitation PSF at $(-x_0, 0)$ and 1D integrated emission PSF at $(0, y)$. Assuming $P_{ex}$ and $P_{em}$ are unimodal distributions with widths $\sigma_{ex} < \sigma_{em}$, the reconstructed PSF $I^x_{PSF}$ is "elliptical" with higher resolution in x-direction. The same process can be applied to y-direction as well. The corresponding equation of Eq. 5 but in the orthogonal direction is

$$I^y_{PSF}(x; x_0, y_0) = P_{ex}(0, -y_0) \tilde{P}_{em}(x, 0; x_0, y_0). \qquad (6)$$

Repeat this process to each scanning location along scanning direction. The sum of these images generates a reconstructed image covering the whole FOV. To recover a more accurate image, we Fourier transform the two images to the frequency domain labeled as $\tilde{F}_x$ and $\tilde{F}_y$. Because the previous reconstruction is along one-dimension (Fig. S1 (a-b)), $\tilde{F}_x$ and $\tilde{F}_y$ are ellipses with long axis in x and y direction respectively, resulting in an anisotropic coverage of frequency domain (Fig. S1 (d-e)). Before normalization, $\tilde{F}_x$ and $\tilde{F}_y$ have different peak values (Fig. S1 (d) is dimmer than Fig. S1 (e)) influenced by alignment. To generate a circular symmetric coverage of Fourier domain (Fig. S1(f)), $\tilde{F}_x$ and $\tilde{F}_y$ were normalized by different weights before overlapping. Median filter to remove random noise and edge enhancement filter to sharpen fine structures were also applied to $\tilde{F}_x$ and $\tilde{F}_y$. The overlay of reconstructed PSF of two orthogonal directions significantly extends the bandwidth coverage along the $k_x$ and $k_y$ axis, beyond the limits imposed by the emission PSF. The inverse Fourier transform of the overlapped frequency components generates the final reconstructed image (Fig. S1 (c) and Fig. 1b).

## 2. mosTF setup and imaging parameters

The mosTF system is based on a standard lineTF design (Fig. 1a). The laser generates femtosecond pulses at a wavelength of 1035 nm (repetition rate 1 MHz, spectral width 10 nm, Monaco, Coherent Inc., CA, USA). The scanning mirror (6350, Cambridge Technology, MA, USA) mechanically scans the beam along the x-axis. The cylindrical lens (*f* = 150 mm) focuses the beam into a line on the grating (20RG1200-1000-2, Newport Co., CA, USA, 1200 grooves/mm). The incident angle θ$_i$ is about 73⁰, so the 1st order diffraction angle is about 17⁰. The grating generates dispersion along the y-axis. L3 (*f* = 300 mm) and L4 (*f* = 75 mm) are relay lenses. A diffractive optical element (MS-635-J-Y-S, HOLO/OR Ltd., Israel) is placed on the conjugate Fourier plane after L3 to generate four parallel lines. The number of parallel lines is calculated from the thermal damage threshold of *in vivo* imaging and the power required for efficient two-photon excitation of a single diffractive spot. After L4, flip mirrors (M1-M3) are placed before L5 to rotate the scanning direction by 90⁰ for the other scanning direction. L5 (*f* = 300 mm) is the tube lens. On the back focal plane, the beam size is about 20 x 20 mm. We overfilled the back aperture so that the excitation PSF of mosTF microscopy is comparable with TPLSM. The FOV is 205 x 205 µm$^2$. L6 ($f$ = 350 mm) is the tube lens in the detection path. The system magnification is about 40x according to the objective magnification and the focal length of the tube lenses. The image is detected by an EMCCD camera (HNu 512, Nuvu Cameras, Canada).

### 3. Image collection for mosTF and lineTF

In the experiments, the EMCCD has 512 x 512 pixels corresponding to a 205 x 205 µm$^2$ FOV. To avoid cross-talk between lines but maximize the number of parallel lines used, we chose the number of parallel lines according to the scattered emission PSF. The scattered emission PSF depends on both the mean-free-path of mouse brain[22-23] and imaging depth. For our *in vivo* experiment, we parallelized four lines with 51 µm separation. Therefore, each line needs to scan a region of 128 x 512 pixels. The PSF of the system is 0.85 µm in lateral direction and 1.39 µm in axial direction (Fig. 2b-d). To fulfill the Nyquist requirement, the pixel size is 0.4 µm. An intermediate image is taken at the position of every pixel. So, the total number of intermediate images is 128 per plane. Based on our previous experiments of single line-scan temporal focusing *in vivo*[5], the exposure time required to collect sufficient emission from one imaging plane is 1.6 s. Thus, for four-line scanning, the exposure time required to collect sufficient emission from one plane is 400 ms. During the 400 ms, we need to capture 128 intermediate images. The desired exposure time of each intermediate image is therefore 3 ms. However, the maximum imaging speed of the EMCCD in our experiment is 63 fps (16 ms), which is the current limitation on mosTF imaging speed. Thus, the exposure time of each intermediate image was set to be 16 ms, and the total exposure time per 2D plane was 2 s (16 ms x 128). Afterwards, scanning was repeated along the orthogonal direction, doubling the total imaging time for a total exposure time of 4 s per 2D plane (16 ms x 128 x 2). When compared to the speed of TPLSM[3-5], which takes about 30 s to collect sufficient emissions from one plane in our *in vivo* preparation, mosTF shows an eight-fold improvement in imaging speed. Noticed that time comparison between different systems are based on collecting similar number of emission photons. Without the speed limitation imposed by the maximum frame rate of the camera, the theoretical speed improvement of mosTF using our 4-line scanning strategy would be approximately 40-fold compared to TPLSM.

mosTF requires a high frame rate camera with low readout noise. As shown in the Theory, the final image is reconstructed by summing the scattered emission photons on adjacent pixels back to the focal point. The readout noise of the camera should be low enough so that summing all these pixels still maintains single photon sensitivity in the final image. At the same time, the overall imaging time of a single plane depends on the maximum frame rate when the sample is bright enough. The EMCCD (HNu 512, Nuvu Cameras, Canada) used in mosTF has 0.05e$^-$ readout noise per frame at the maximum gain. The camera was tested by imaging a blank cover glass to measure the background intensity. We mimicked the lineTF image by directly summing 128 intermediate images without reconstruction. In this case, the background is 2.57e- (±2.14e-), which is similar to a standard lineTF image. After mosTF reconstruction, the background is 0.13e- (±0.39e-) (Fig. S2). If imaging brighter fluorescence samples, such as soma of neuron instead of spines, the exposure time per pixel could be much shorter to achieve sufficient SNR. Recent development of high frame rate intensified sCMOS camera can significantly alleviate camera speed limitation. An alternative approach is to image in a de-scanned geometry. In this approach, each scan line and the associated scanned region are projected to a column at the camera using elliptical optics. Instead of reading out the whole camera, approximately a column for each scan line needs to be readout. In the case of four scan lines, the effective frame rate using the intensified sCMOS camera can exceed 4 kHz. For 512x512 image, final maximum frame rate can reach 31 Hz while providing excellent SNR image with over 250 μs pixel dwell time.

## 4. Image collection for TPLSM

TPLSM was performed using a custom-built microscope with a Mai Tai HP Ti: Sapphire laser (Spectra Physics) tuned to 1030 nm as the source of excitation. The power delivered to the specimen ranged from 30 to 50 mW depending on imaging depth. Galvanometric XY scanning mirrors (6215H, Cambridge Technology) and a piezo actuator Z positioning system (Piezosystem Jena) were used for XY and Z movement, respectively. The pixel size was 250 nm in XY, and the Z step size was 1 μm. The dwell time per pixel was 40 μs. The dwell time was determined based on our previous data as the time required to collect sufficient emission photons from the dendrites of L2/3 pyramidal neurons expressing an eYFP cell fill *in vivo*[3,4]. The beam was focused by a 20x/1.0 NA water immersion objective lens (W Plan-Apochromat, Zeiss). Emissions were collected by the same objective lens, passed through an IR blocking filter (E700SP, Chroma Technology), and separated by dichroic mirrors at 520 nm and 560 nm. After passing through three independent band-pass filters (485/70 nm, 550/100 nm, and 605/75 nm), emissions were collected simultaneously onto three separate PMTs. Raw 2-photon scanning data were processed for spectral linear unmixing and converted into a tif Z stack using Matlab (Mathworks) and ImageJ (NIH). Only the fluorescence assigned to the 550/100 nm (i.e. yellow) channel was used for this study to visualize the dendritic morphology of L2/3 pyramidal neurons expressing an eYFP cell fill.

## 5. Bead sample preparation for scattering and non-scattering imaging conditions

In addition to *in vivo* specimens, bead samples with known scattering mean free paths were prepared to evaluate the performance of this imaging system. The mean free path length ($l_s^{em}$) of emission photons was calculated by $l_s^{em} = 1/\mu_s$, where $\mu_s$ is the scattering coefficient. For *in vivo* imaging, the brain tissue and blood are a highly heterogeneous. The scattering coefficient of mouse or rat brain varies slightly in the literature[23-24]. The mean free path length $l_s^{em}$ at 532 nm is about 43.5-58.8 μm calculated from the reported scattering parameters measured *in vitro*[22-23]. These values informs our choice of intralipid concentration and thickness to mimic brain scattering. For a 2% intralipid solution, $\mu_s$ equals 16 mm$^{-1}$ according to literature[25]. Since the scattering coefficient is linearly proportional to concentration, $l_s^{em}$ of

0.2% intralipid solution is 0.625 mm. Our samples consisted of 200 nm diameter red fluorescent beads (the same as used for the PSF measurement) mounted in clear medium (Fluoromount-G®, SouthernBiotech, AL, USA). The specimens were imaged through 2 mm of water as controls. The same specimen was then imaged through 2 mm of 0.2% intralipid solution as scattering phantoms. These phantoms have scattering equivalent to $3.2 l_s^{em}$ in the brain or corresponding to a tissue depth of roughly 160 μm. Finally, the 200 nm beads mimic the synapses (about 1 μm).

## 6. *In vivo* specimen preparation

All animal procedures were approved by the Massachusetts Institute of Technology Committee on Animal Care and meet the NIH guidelines for the care and use of vertebrate animals. *In utero* electroporation was performed on embryonic day 15.5 timed pregnant C57BL/6J mice to label a sparse population of L2/3 pyramidal neurons with an eYFP cell fill, enabling visualization of the neurons' dendritic morphology. Constructs used for *in utero* electroporation were a cre-dependent eYFP cell fill (*pFUdioeYFPW*[3]) and a *Cre* plasmid[26] at concentrations of 0.7 μg/μl and 0.03 μg/μl, respectively, with 0.1% Fast Green for visualization. A total of 0.5-1.0 μl of the plasmid solution was injected into the right lateral ventricle, and five pulses of 36V (duration 50 ms, frequency 1 Hz) targeting the visual cortex were delivered from a square-wave electroporator (ECM830, Harvard Apparatus). Pups were then reared to adolescence (P44) and implanted with a 5 mm cranial window over the right hemisphere as described previously[27]. After 2 weeks of surgery recovery, animals were fitted with a custom head mount to enable fixation to the microscope stage. All imaging took place under isoflurane anesthesia (1.25%) with the head mount fixed to the microscope stage. The animal whose data are shown in Fig. 4 was imaged at postnatal day (P) 63.

**Acknowledgments**

This work was supported by National Institutes of Health (NIH) funding 5-P41-EB015871-27 (P. T. C. S.), Hamamatsu Corporation; Samsung Advanced Institute of Technology; Singapore–Massachusetts Institute of Technology Alliance for Research and Technology (SMART) Center, BioSystems and Micromechanics (BioSyM); NIH funding 5R21NS091982-02, 1-U01-NS090438-01, and 1-R21-NS105070-01 (E. N. and P. T. C. S.); F32 MH115441 (J. B.) and Jeffry M. and Barbara Picower Foundation (E. N.); Center for Advanced Imaging at Harvard University (D. W.).

Supplementary figure 1

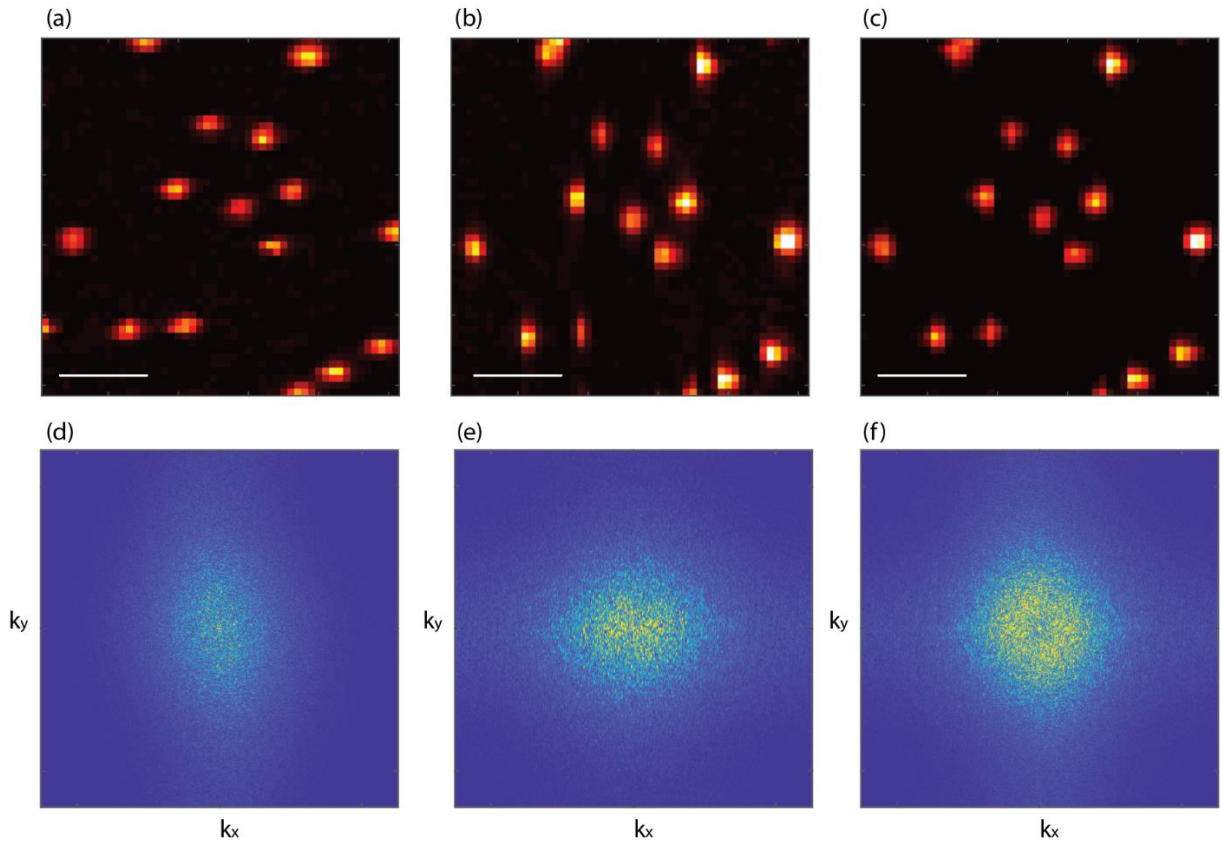

Figure S1. Orthogonal scanning provides an isotropic PSF. (a) Reconstructed image of 200 nm beads from horizontal scanning only. (b) Reconstructed image of 200 nm beads from vertical scanning only. (c) Combination of (a) and (b) to generate isotropic PSF. Scale bar of (a-c), 5 μm. Figures (a-c) are normalized to the same color scale. (d) Fourier transform of (a). The frequency coverage after reconstruction in the horizontal direction is extended in the vertical direction. (e) Fourier transform of (b). (f) Combination of (d) and (e) with a weighting factor to balance the frequency coverage in two directions. The frequency coverage is isotropic in the Fourier domain. The inverse Fourier transform of (f) generates (c).

Supplementary figure 2

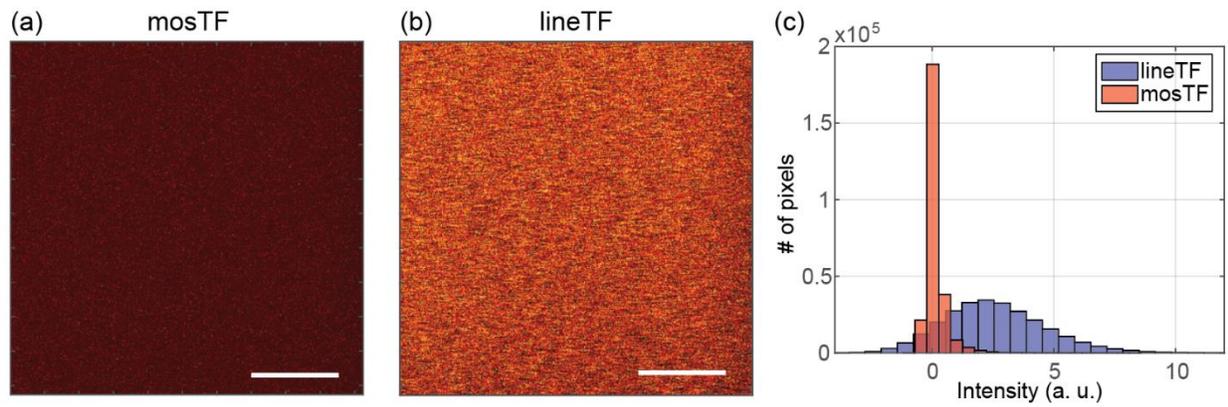

Figure S2. Background measurement of (a) mosTF and (b) lineTF. These two figures are normalized to the same color scale [-1 9]. Scale bar, 50 µm. (c) Pixel value histogram of mosTF and lineTF. The camera pixel has an offset of 300 counts to avoid negative value of intensity, which was subtracted during our calculation. The mean value of mosTF background is 0.13±0.39$e^-$, and the mean value of lineTF background is 2.57±2.14$e^-$.